\documentclass[aps,twocolumn,showpacs,floatfix]{revtex4}
\usepackage{amsmath}
\usepackage{amsfonts}
\usepackage{amssymb}
\usepackage{graphics}
\usepackage[all]{xy}
\begin{document}
\title{Photon-Photon Correlations as a Probe of Vacuum Induced Coherence Effects}
\author{Sumanta Das and G. S. Agarwal}
\affiliation{
Department of Physics, Oklahoma State University, Stillwater,
Oklahoma 74078, USA}
\date{\today}
\begin{abstract}
We present new experimental implications of the effects of vacuum
induced coherence  on the photon -photon correlation in the
$\pi$-polarized fluorescence in j = 1/2 to j = 1/2 transition.
These effects should be thus observable in measurements of photon
statistics in for example Hg and Ba ion traps.
\end{abstract}
\maketitle

\section{Introduction}
An early work \cite{gs} had predicted a very unusual effect of
quantum interference in the problem of spontaneous emission. It
was for example shown that in a degenerate V-system one could get
population trapping and generation of quantum coherences in the
excited states. One of the key conditions for the occurrence was
that the dipole matrix elements of the two transitions from the
excited states were orthogonal. The later condition is difficult
to meet though very large body of theoretical literature has been
devoted to the subject of vacuum induced coherences
\cite{keitel,paspalaski, zhou,menon,javen}. It was also suggested
how the above condition on dipole matrix elements can be bypassed
if we consider anisotropic vacuum \cite{gsa}  which for example
would be the case while considering emission from excited atoms on
nano particles \cite{adel}. It is clearly important to find out
easily realizable systems so that experimental results can be
obtained. Kiffner. et. al.\cite{evers} showed that one very
important case would involve j = 1/2 to j = 1/2 transition. They
calculated the spectrum of the emitted radiation and showed how
vacuum induced coherences change the spectrum of the emitted
radiation. There are many systems where it is easy to find j = 1/2
to j = 1/2 transition. In fact in an early experiment of Eichmann
et. al.\cite{eich} such transitions in $^{198}Hg^{+}$ were used to
examine interferences in a system of two ions. A recent work uses
the j = 1/2 to j = 1/2 transition in $^{138} Ba^{+}$ ions
\cite{blatt}. While the results of Kiffner. et. al. on the
spectrum are quite interesting the current experimental efforts
\cite{blatt,kuch,monroe} are more focussed on the study of
photon-photon correlations. Thus an important question would be --
do the vacuum induced coherences significantly affect the
photon-photon correlations ? This is the question we answer in
affirmative. \\
It may be added that the photon-photon correlations have acquired
new significance in the context of quantum information processing
and quantum imaging as well as in interferences from independent
atoms \cite{kuch,monroe,thiel}. Thus it is important to have a
consistent calculation of such correlations in situations where
vacuum induced coherence (VIC) effects are important.\\
The organization of this paper is as follows-In Sec 2 we introduce
the model and present the working equations. In Sec 3 we calculate
the photon-photon correlations both in presence and in absence of
the vacuum induced interference effects. In Sec 4 we present
numerical results to highlight the effects of vacuum induced
coherences on photon-photon correlations. In Sec 5 we conclude
with the outlook and future directions.\\

\section{Model}
\begin{figure}[!h]
\begin{center}
\scalebox{0.65}{\includegraphics{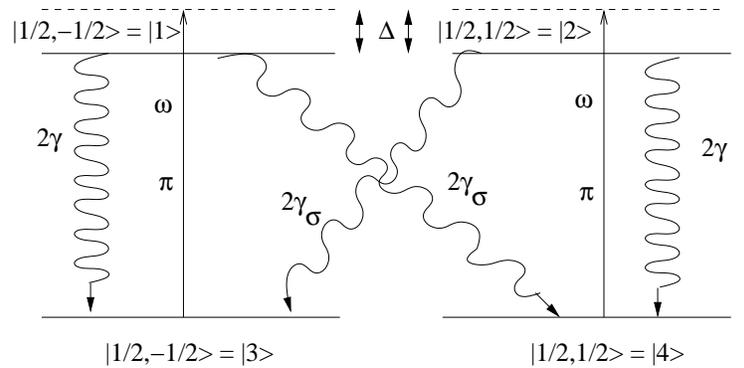}} \caption{Schematic
diagram of a four-Level atom modelled by j = 1/2 to j = 1/2
transition}
\end{center}
\end{figure}
 The Fig.1. shows the level scheme of a four-level atom modelled by j = 1/2 to j = 1/2 transition.
 This kind of level scheme is easily realizable and has already been studied, for example
in  $^{198}$Hg$^+$ \cite{eich} and $^{138}$Ba$^+$\cite{blatt}
ions. The ground level is $6s^{2}S_{1/2}$ and the excited level is
$6p^{2}P_{1/2}$. Each of these levels are two fold degenerate. The
transitions $|1\rangle\leftrightarrow|4\rangle$ and
$|2\rangle\leftrightarrow|3\rangle$ couple to $\sigma^{+}$ and
$\sigma^{-}$ polarized light respectively. The transitions
$|1\rangle\leftrightarrow|3\rangle$ and
$|2\rangle\leftrightarrow|4\rangle$ couple to light linearly
polarized along the $\bf{e}$$_z $ and their dipole moments are
antiparallel. The spontaneous decays of the excited state to the
two ground states are given by $2\gamma$ and $2\gamma_{\sigma}$ as
shown in the figure. The electric  dipole moment operator  for
this level scheme is defined as
\begin {eqnarray}
\mathbf{d} & = &\sum_{ij}
\mathbf{d}_{ij}A_{ij}\;,\nonumber\\
A_{ij} & = &|i\rangle\langle j|.\qquad (i,j = 1,..,4)\;
\end{eqnarray}
The non-vanishing matrix elements of the electric dipole moment
operator $\mathbf{d}$ can be found using the Wigner-Eckart theorem
and are given by,
\begin{eqnarray}
\vec{d}_{31}=- \vec{d}_{42} = -\frac{1}{\sqrt{6}}\mathcal{D}\hat{e}_z\;,\nonumber\\
\label{d}\vec{d}_{41} =  \vec{d}^{\ast}_{32} =
\frac{1}{\sqrt{3}}\mathcal{D}\hat{\epsilon}_-\;,
\end{eqnarray}
with $\hat{\epsilon}_-$ = $(\hat{x}-i\hat{y})/\sqrt{2}$. In Eq.
(2) $\mathcal{D}$ denotes the reduced matrix element of the dipole
moment operator $\mathbf{d}$. The four-level system is driven by a
$\pi$ polarized monochromatic field of frequency $\omega$ ,
\begin{equation}
\mathbf{E}(t) =\vec{ \mathcal{E}_0}e^{-i\omega t}\mathbf{e}_z +
c.c.\; ,
\end{equation}
were c.c is the complex conjugate. With this particular choice of
polarization, the driving field couples only to the two
antiparallel dipole moments $\vec{d}_{31}$ and $\vec{d}_{42}$. The
total Hamiltonian for this atom-field system is then given by
\begin{equation}
\mathcal{H} = \mathcal{H}_{A}+\mathcal{H}_{I}\;,
\end{equation}
where the unperturbed Hamiltonian for the atom is,
\begin{equation}
\mathcal{H}_{A} = \hbar\sum_{i=1}^{4}\omega_{i}|i\rangle\langle
i|\;,
\end{equation}
an the interaction Hamiltonian for this system is given by
\begin{eqnarray}
\label{rho}\mathcal{H}_{I} & = &-\mathbf{d}.\mathbf{E(t)}\nonumber\\
& = &\hbar\Omega_{c}(|1\rangle\langle 3| -|2\rangle\langle
4|)e^{-i\omega t} + H.c\;,
\end{eqnarray}
where H.c is the Hermitian conjugate and $\Omega_c$ is the Rabi
frequency defined by
\begin{equation}
\Omega_{c} =\frac{ \vec{d}_{42}\cdot\vec{\mathcal{E}_0}}{\hbar}\;,
\end{equation}
The time evolution of this four level system is investigated by studying the density matrix equation.
The spontaneous emission is included via the master equation techniques. Following the standard procedure \cite{gs} we obtain,
\begin{eqnarray}
\dot{\rho}& = &-\frac{i}{\hbar}\lbrack\mathcal{\hat{H}}, \rho\rbrack+\mathcal{L}\rho\;,\nonumber\\
\mathcal{L}\rho & = &-\gamma_{\sigma}\lbrack|1\rangle\langle1|\rho+|2\rangle\langle 2|\rho+\rho|1\rangle\langle 1|\nonumber\\
& + &  \rho|2\rangle\langle 2|-2|3\rangle\langle3|\rho_{22}-2|4\rangle\langle4|\rho_{11} \rbrack\nonumber\\
& - & \gamma\lbrack|1\rangle\langle1|\rho+|2\rangle\langle 2|\rho+\rho|1\rangle\langle 1| \nonumber\\
& + &  \rho|2\rangle\langle
2|-2|3\rangle\langle3|\rho_{11}-2|4\rangle\langle4|\rho_{22}\rbrack
\nonumber\\
& + & \gamma
\lbrack|4\rangle\langle3|\rho_{21}+|3\rangle\langle4|\rho_{12}\rbrack\;,
\end{eqnarray}
The last two terms in Eq. (8) arise from the vacuum induced
interference and it comes as the dipole matrix elements
$\vec{d}_{13}$ and $\vec{d}_{24}$ are anti-parallel. In a frame
rotating with the frequency of the coherent drive the density
matrix equations are,
\begin{eqnarray}
\label{den}\dot{\tilde{\rho}}_{11} & = &i\Omega^{\ast}_{c}\tilde{\rho}_{13}-i\Omega_{c}\tilde{\rho}_{31}-2\Gamma\tilde{\rho}_{11}\;,\nonumber\\
\dot{\tilde{\rho}}_{22} & = &i\Omega_{c}\tilde{\rho}_{42}-i\Omega^{\ast}_{c}\tilde{\rho}_{24}-2\Gamma\tilde{\rho}_{22}\;,\nonumber\\
\dot{\tilde{\rho}}_{33} & = &i\Omega_{c}\tilde{\rho}_{31}-i\Omega^{\ast}_{c}\tilde{\rho}_{13}+\gamma_{\sigma}\tilde{\rho}_{22}+\gamma\tilde{\rho}_{11}\;,\nonumber\\
\dot{\tilde{\rho}}_{12} & = &-i\Omega_{c}\tilde{\rho}_{32}-i\Omega^{\ast}_{c}\tilde{\rho}_{14}-2\Gamma\tilde{\rho}_{12}\;,\nonumber\\
\dot{\tilde{\rho}}_{13} & = &-i\Delta\tilde{\rho}_{13}+i\Omega_{c}(\tilde{\rho}_{11}-\tilde{\rho}_{33})-\Gamma\tilde{\rho}_{13}\;,\\
\dot{\tilde{\rho}}_{14} & = &-i\Delta\tilde{\rho}_{14}-i\Omega_{c}\tilde{\rho}_{12}-i\Omega_{c}\tilde{\rho}_{34}-\Gamma\tilde{\rho}_{14}\;,\nonumber\\
\dot{\tilde{\rho}}_{23} & = &-i\Delta\tilde{\rho}_{23}+i\Omega_{c}\tilde{\rho}_{21}+i\Omega_{c}\tilde{\rho}_{43}-\Gamma\tilde{\rho}_{23}\;,\nonumber\\
\dot{\tilde{\rho}}_{24} & = &-i\Delta\tilde{\rho}_{24}-i\Omega_{c}(\tilde{\rho}_{22}-\tilde{\rho}_{44})-\Gamma\tilde{\rho}_{24}\;,\nonumber\\
\dot{\tilde{\rho}}_{34} & =
&-i\Omega_{c}\tilde{\rho}_{32}-i\Omega^{\ast}_{c}\tilde{\rho}_{14}-\gamma
q\tilde{\rho}_{12}\;,\nonumber\\\nonumber
\end{eqnarray}
where
\begin{eqnarray}
\tilde{\rho}_{ii} = \rho_{ii}\qquad ;\qquad \tilde{\rho}_{ij} =
\rho_{ij}e^{-i\omega t}\;\nonumber\\
 \Gamma = (\gamma_{\sigma} +
\gamma) ; \qquad\Delta = \omega-\omega_{13} =
\omega-\omega_{24}\;,
\end{eqnarray}
The remaining equations can be generated by taking complex
conjugates and using Tr$\{\rho\} = 1$. The steady state solution
of Eq. (10) are found to be

\begin{eqnarray}
\tilde{\rho}_{12} & = &\tilde{\rho}_{14}=\tilde{\rho}_{32}=\tilde{\rho}_{34} = 0\;,\\
\nonumber\\
\label{sol}\tilde{\rho}_{11}& = &\tilde{\rho}_{22}=\frac{1}{2}\frac{|\Omega_{c}|^2}{\lbrack2|\Omega_{c}|^2+\Gamma^2+\Delta^2\rbrack}\;,\nonumber\\
\tilde{\rho}_{33}& = &\tilde{\rho}_{44}=\frac{1}{2}\frac{|\Omega_{c}|^2+\Gamma^2+\Delta^2}{\lbrack2|\Omega_{c}|^2+\Gamma^2+\Delta^2\rbrack}\;,\\
\tilde{\rho}_{13}& = &-\tilde{\rho}_{24} =
-\frac{i\Omega_{c}}{\Gamma+i\Delta}\{\frac{1}{2}\frac{\Gamma^2+\Delta^2}{\lbrack2|\Omega_{c}|^2+\Gamma^2+\Delta^2\rbrack}\}\;,\nonumber\\\nonumber
\end{eqnarray}
 As can be seen from Eqs. (11) and (12)
the vacuum induced interference has no effect on the steady state
solutions. Clearly vacuum induced coherence can show up in
dynamical quantities.

\section{Photon-Photon Correlations}
Since the objective of this paper is to investigate the observable
consequences of the vacuum induced coherence ; we focus our
attention on the photon statistics of the radiation emitted by our
model system. We in particular will calculate photon-photon
 correlations as currently considerable experimental effort is on such correlations.
  For this we need to know how to relate the atomic properties
 with the statistical properties of the spontaneously emitted radiation. The answer to this question
 already exists in quantum theory. In fact from the existing \cite{gsab} literature, we know that the
 positive frequency part of the electric-field operator at a point $\vec{r}$
 in the far-field zone can be written in terms of the atomic operators as
\begin{eqnarray}
\label{field}\mathbf{E}^{+}(\vec{r},t) & = &
\mathbf{E}^{+}_{0}(\vec{r},t)-k^2_{0}\sum_{i}
\{\lbrack\hat{R}_{i}\times(\hat{R}_{i}\times\vec{d}_{31})A_{31}\rbrack\nonumber\\
&+&\lbrack\hat{R}_{i}\times(\hat{R}_{i}\times\vec{d}_{42})A_{42}\rbrack\nonumber\\
&+&\lbrack\hat{R}_{i}\times(\hat{R}_{i}\times\vec{d}_{32})A_{32}\rbrack\nonumber\\
&+&\lbrack\hat{R}_{i}\times(\hat{R}_{i}\times\vec{d}_{41})A_{41}\rbrack\}R^{-1}_{i} \nonumber\\
&\times& e^{-i(k_{0}\hat{r}\cdot\vec{r}_{i}+\omega\tau)}\;,
\end{eqnarray}
where $\vec{R}_{i} = \vec{r} -\vec{r}_{i}$ , $\vec{r}$ being the
distance of the point of observation from the origin  and
$\vec{r}_{i}$ being the position of the atom from the origin.
Further $\tau = t-\frac{r}{c}$ is the retarded time, $ k_{o} =
\frac{\omega_{0}}{c}$, $\omega_{0} = \omega_{13} = \omega_{24}$,
$\vec{d}_{ij}$ is the electric dipole moment operator and the
atomic operators are as defined in Eq. (1).
 The first term on the right of  Eq. (\ref{field})
 is the free field term and the second term is the retarded dipole field emitted by the atom.
The emitted radiation consist of different polarization
components-- the $\pi$ and the $\sigma$ polarized components. The
terms $A_{31}$ and $A_{42}$ correspond to $\pi$ polarization
whereas the ones $A_{32}$ and $A_{41}$ correspond to $\sigma$
polarization. We next calculate the photon-photon correlations and
the normalized second order correlations for the
 emitted radiations from the $\pi$ transitions of this driven four-level atom.
 For $\pi$ polarization the relevant part of the electric field operator is
 given by,
\begin{eqnarray}
\label{fil}\mathbf{E}^{+}(\vec{r},t) & = &\mathbf{E}^{+}_{0}(\vec{r},t )-(\frac{\omega_{0}}{c})^{2}
\frac{1}{r}([\hat{n}\times(\hat{n}\times\vec{d}_{31})] {|3\rangle\langle 1|}_{\tau}\nonumber\\
&&+[\hat{n}\times(\hat{n}\times\vec{d}_{42})] {|4\rangle\langle
2|}_{\tau})\;,
\end{eqnarray}
In the lowest order correlation the free field term of Eq.
(\ref{fil}) does not contribute. This can be seen directly from
the definition of quantized fields \cite{gsab}, the fact that the
field is initially in the vacuum state and the expression for the
normally ordered correlation function for the field,
 $\langle\mathbf{E}^{-}(\vec{r},t)\mathbf{E}^{+}(\vec{r'},t')\rangle$.
 Hence with no contribution from the free field term
the intensity $I_{\pi}$ of the light emitted on the $\pi$
transition from the atom is ,
\begin{eqnarray}
\label{intp}\langle I_{\pi}\rangle& = & \langle\mathbf{E}^{-}_{\pi}(\vec{r},t)\cdot\mathbf{E}^{+}_{\pi}(\vec{r},t)\rangle\nonumber\\
& = &(\frac{\omega_{0}}{c})^{4}\frac{1}{r^2}\langle[\hat{n}\times(\hat{n}\times\vec{d}_{31})]^{\ast}\cdot[\hat{n}\times(\hat{n}\times\vec{d}_{31})]{|1\rangle\langle 1|}_{\tau}\nonumber\\
&&
+[\hat{n}\times(\hat{n}\times\vec{d}_{42})]^{\ast}\cdot[\hat{n}\times(\hat{n}\times\vec{d}_{42})]{|2\rangle\langle
2|}_{\tau}\rangle\;,
\end{eqnarray}
where we have taken our origin at the location of the atom ,
$\vec{r} = \hat{n}r$, $\tau$ is the retarded time and we used the
property $A_{ij}A_{kl} = A_{il}\delta_{kj}$. The negative
frequency part of the electric field operator
$\mathbf{E}^{-}(\vec{r},t)$ can be found by taking the complex
conjugate of the positive frequency part. Now if we assume that
the point of observation lies perpendicular to both the
polarization and propagation direction we have from Eq.
(\ref{intp})
\begin{equation}
\langle I_{\pi}\rangle
 = (\frac{\omega_{0}}{c})^{4}\frac{1}{r^2}(|\vec{d}_{31}|^{2}\langle|1\rangle\langle 1|\rangle _{\tau}+
 |\vec{d}_{42}|^{2}\langle|2\rangle\langle 2|\rangle_{\tau})\;,
\end{equation}
Eq.(16) can be further simplified using Eqs. (\ref{d}) and
(\ref{sol}), where in using Eq. (\ref{sol}) we have assumed that
observation is been made at long time limit. The final expression
for $I_{\pi}$ in the long time limit (steady state) is then,
\begin{equation}
\langle I_{\pi}\rangle^{st} =
(\frac{\omega_{0}}{c})^{4}\frac{|\mathcal{D}|^{2}}{6r^2}\frac{|\Omega_{c}|^2}
{\lbrack2|\Omega_{c}|^2+\Gamma^2+\Delta^2\rbrack}\;,
\end{equation}
Eq.(17) clearly show that intensity emitted on the $\pi$
transitions is not altered by vacuum induced coherences and is
simply proportional to the steady state population of the
excited states.\\
 Let us now investigate what happens incase of
two time photon-photon correlations on the $\pi$ transitions. The
two-time photon-photon correlation for the level scheme in Fig.1
can be written as
\begin{eqnarray}
\langle I_{\pi}(t+\tau)I_{\pi}(t)\rangle & = & \langle\mathbf{E}^{-}_{\pi}(\vec{r},t)\mathbf{E}^{-}_{\pi}(\vec{r},t+\tau):\nonumber\\
&&\mathbf{E}^{+}_{\pi}(\vec{r},t+\tau)\mathbf{E}^{+}_{\pi}(\vec{r},t)\rangle\nonumber\\
& = &(\frac{\omega_{0}}{c})^{8}\frac{1}{r^4}\{[\hat{n}\times(\hat{n}\times\vec{d}_{31})]^{\ast}\nonumber\\
&&\cdot[\hat{n}\times(\hat{n}\times\vec{d}_{31})]\}^{2}\nonumber\\
&&\langle(|1\rangle\langle 3|-|2\rangle\langle 4|)_{t}(|1\rangle\langle 1|+|2\rangle\langle 2|)_{t+\tau}\nonumber\\
&& (|3\rangle\langle 1|-|4\rangle\langle 2|)_t\rangle\;,
\end{eqnarray}
The two-time correlation function which appears in Eq. (18) is to
be obtained from the solution of the time-dependent density matrix
equations (Eq.(9)) and the quantum regression theorem \cite{lax}.
A closer look at Eq. (9) show that eight of the fifteen equations
form a closed set of linear equations which can be solved to find
$|1\rangle\langle1|_{t+\tau}, |2\rangle\langle2|_{t+\tau}$ and
hence the term $(|1\rangle\langle1|+ |2\rangle\langle2|)_{t+\tau}$
in Eq. (18). Before going further let us list those eight
equations,
\begin{eqnarray}
\label{coup}\dot{\tilde{\rho}}_{11} & = &i\Omega^{\ast}_{c}\tilde{\rho}_{13}-i\Omega_{c}\tilde{\rho}_{31}-2\Gamma\tilde{\rho}_{11}\;,\nonumber\\
\dot{\tilde{\rho}}_{33} & = &i\Omega_{c}\tilde{\rho}_{31}-i\Omega^{\ast}_{c}\tilde{\rho}_{13}+\gamma_{\sigma}\tilde{\rho}_{22}+\gamma\tilde{\rho}_{11}\;,\nonumber\\
\dot{\tilde{\rho}}_{13} & = &-i\Delta\tilde{\rho}_{13}+i\Omega_{c}(\tilde{\rho}_{11}-\tilde{\rho}_{33})-\Gamma\tilde{\rho}_{13}\;,\nonumber\\
\dot{\tilde{\rho}}_{31} & = &i\Delta\tilde{\rho}_{31}-i\Omega^{\ast}_{c}(\tilde{\rho}_{11}-\tilde{\rho}_{33})-\Gamma\tilde{\rho}_{31}\;,\\
\dot{\tilde{\rho}}_{22} & = &i\Omega_{c}\tilde{\rho}_{42}-i\Omega^{\ast}_{c}\tilde{\rho}_{24}-2\Gamma\tilde{\rho}_{22}\;,\nonumber\\
\dot{\tilde{\rho}}_{44} & = &i\Omega^{\ast}_{c}\tilde{\rho}_{24}-i\Omega_{c}\tilde{\rho}_{42}+\gamma_{\sigma}\tilde{\rho}_{11}+\gamma\tilde{\rho}_{22}\;,\nonumber\\
\dot{\tilde{\rho}}_{24} & = &-i\Delta\tilde{\rho}_{24}-i\Omega_{c}(\tilde{\rho}_{22}-\tilde{\rho}_{44})-\Gamma\tilde{\rho}_{24}\;,\nonumber\\
\dot{\tilde{\rho}}_{42} & =
&i\Delta\tilde{\rho}_{42}+i\Omega^{\ast}_{c}(\tilde{\rho}_{22}-\tilde{\rho}_{44})-\Gamma\tilde{\rho}_{42}\;,\nonumber\nonumber
\end{eqnarray}
In compact notation this equations can be written as,
\begin{equation}
\dot{\tilde{\rho}} = \mathcal{M}\tilde{\rho}\;,
\end{equation}
where $\dot{\tilde{\rho}}$ , $\tilde{\rho}$ are (8$\times$1)
column matrix and $\mathcal{M}$ is a (8$\times$8) square matrix.
Now using the method depicted in \cite{gsap} and using
Eq.(\ref{coup}) the solution of
$\langle|1\rangle\langle1|_{t+\tau}\rangle$ and
$\langle|2\rangle\langle2|_{t+\tau}\rangle$ can be expressed in
the form
\begin{eqnarray}
\langle|1\rangle\langle1|_{t+\tau}\rangle & = &f_{11}(\tau)\langle|1\rangle\langle1|_{t}\rangle+f_{12}(\tau)\langle|3\rangle\langle|3|_{t}\rangle\nonumber\\
&+&f_{13}(\tau)\langle|3\rangle\langle1|_{t}\rangle+f_{14}(\tau)\langle|1\rangle\langle3|_{t}\rangle\nonumber\\
&+&f_{15}(\tau)\langle|2\rangle\langle2|_{t}\rangle+f_{16}(\tau)\langle|4\rangle\langle4|_{t}\rangle\nonumber\\
&+&f_{17}(\tau)\langle|4\rangle\langle2|_{t}\rangle+f_{18}(\tau)\langle|2\rangle\langle4|_{t}\rangle\;,
\end{eqnarray}
\begin{eqnarray}
\langle|2\rangle\langle2|_{t+\tau}\rangle & = &f_{51}(\tau)\langle|1\rangle\langle1|_{t}\rangle+f_{52}(\tau)\langle|3\rangle\langle|3|_{t}\rangle\nonumber\\
&+&f_{53}(\tau)\langle|3\rangle\langle1|_{t}\rangle+f_{54}(\tau)\langle|1\rangle\langle3|_{t}\rangle\nonumber\\
&+&f_{55}(\tau)\langle|2\rangle\langle2|_{t}\rangle+f_{56}(\tau)\langle|4\rangle\langle4|_{t}\rangle\nonumber\\
&+&f_{57}(\tau)\langle|4\rangle\langle2|_{t}\rangle+f_{58}(\tau)\langle|2\rangle\langle4|_{t}\rangle\;,
\end{eqnarray}
where the f's are defined by
\begin{equation}
f_{ik}(\tau) = (e^{\mathcal{M}\tau})_{ik}\;,
\end{equation}
and
\begin{eqnarray}
\mathcal{M}_{ik}  &= &\sum_{l}P_{il}\Lambda_{ll}P^{-1}_{lk}\nonumber\\
(e^{\mathcal{M}\tau})_{ik}& = &
\sum_{l}P_{il}e^{\Lambda_{ll}\tau}P^{-1}_{lk}\;,
\end{eqnarray}
Here we have diagonalized the matrix $\mathcal{M}$  with $\Lambda$
being the eigenvalues and P being the corresponding eigenvectors.
We now make use of the quantum regression theorem to obtain the
two time correlation function as,
\begin{eqnarray}
\langle B^{\dagger}(t)(|1\rangle\langle 1|+|2\rangle\langle
2|)_{t+\tau}B(t)\rangle & = & F_{1}(\tau)\langle
B^{\dagger}(t)|1\rangle\langle1|_{t}B(t)\rangle\nonumber\\
&+&F_{2}(\tau)\langle B^{\dagger}(t)|3\rangle\langle3|_{t}B(t)\rangle\nonumber\\
&+&F_{3}(\tau)\langle B^{\dagger}(t)|3\rangle\langle1|_{t}B(t)\rangle\nonumber\\
&+&F_{4}(\tau)\langle B^{\dagger}(t)|1\rangle\langle3|_{t}B(t)\rangle\nonumber\\
&+&F_{5}(\tau)\langle B^{\dagger}(t)|2\rangle\langle2|_{t}B(t)\rangle\nonumber\\
&+&F_{6}(\tau)\langle B^{\dagger}(t)|4\rangle\langle4|_{t}B(t)\rangle\nonumber\\
&+&F_{7}(\tau)\langle B^{\dagger}(t)|4\rangle\langle2|_{t}B(t)\rangle\nonumber\\
&+&F_{8}(\tau)\langle B^{\dagger}(t)|2\rangle\langle4|_{t}B(t)\rangle\;,\nonumber\\
\end{eqnarray}
where we define the operator B as,
$B^{\dagger}(t)=(|1\rangle\langle3|-|2\rangle\langle4|)_{t}$ ;
$B(t)$ = $(B^{\dagger}(t))^{\dag}$ and $F_{i}(\tau) =
f_{1i}(\tau)+f_{5i}(\tau)$. Using this new definition of the
operator in Eq. (18), the expression for the two-time
photon-photon correlation becomes,
\begin{eqnarray}
\langle I_{\pi}(t+\tau)I_{\pi}(t) \rangle& =
&(\frac{\omega_{0}}{c})^{8}\frac{1}{r^4}
\{[\hat{n}\times(\hat{n}\times\vec{d}_{31})]^{\ast}\nonumber\\
&&\cdot[\hat{n}\times(\hat{n}\times\vec{d}_{31})]\}^{2}\;\\
&\times&\langle B^{\dagger}(t)(|1\rangle\langle
1|+|2\rangle\langle 2|)_{t+\tau}B(t)\rangle\;,\nonumber\nonumber
\end{eqnarray}
which when Eq. (25) is used, simplifies to
\begin{eqnarray}
\langle I_{\pi}(t+\tau)I_{\pi}(t) \rangle& =
&(\frac{\omega_{0}}{c})^{8}\frac{1}{r^4}
\{[\hat{n}\times(\hat{n}\times\vec{d}_{31})]^{\ast}\nonumber\\
&&\cdot[\hat{n}\times(\hat{n}\times\vec{d}_{31})]\}^{2}\;\\
&\times&(F_{2}(\tau)\langle|1\rangle\langle1|\rangle_{t}+F_{6}(\tau)\langle|2\rangle\langle2|\rangle_{t})\;,\nonumber
\\\nonumber
\end{eqnarray}
In the long time limit
$\langle|1\rangle\langle1|\rangle_{t}\equiv\tilde{\rho}_{11}(t)$
and $\langle|2\rangle\langle2|\rangle_{t}
\equiv\tilde{\rho}_{22}(t)$, where $\tilde{\rho}_{11}(t)
,\tilde{\rho}_{22}(t)$ are the steady state populations of the
excited states given by Eq. (\ref{sol}). Now following our
assumption that the point of observation lies perpendicular to
both the polarization and propagation
 directions and substituting for $\tilde{\rho}_{11}$ , $\tilde{\rho}_{22}$ from Eq. (\ref{sol}), Eq. (27) can be
 further simplified.
 The final expression for the two-time photon-photon correlation is then,
\begin{eqnarray}
G_{\pi}^{(2)}(\tau) & = &\langle I_{\pi}(t+\tau)
I_{\pi}(t) \rangle\nonumber\\
&=&(\frac{\omega_{0}}{c})^{8}\frac{|\mathcal{D}|^4}{36r^4}(F_{2}(\tau)+
F_{6}(\tau))\nonumber\\
&\times&(\frac{1}{2}\frac{|\Omega_{c}|^2}{\lbrack2|\Omega_{c}|^2+\Gamma^2+\Delta^2\rbrack})\;,
\end{eqnarray}
where we have used Eq. (2) for the dipole matrix elements. Note
that $F_{2}(\tau)[F_{6}(\tau)]$ is the sum of probabilities of
finding the atom in the states $|1\rangle $ and $|2\rangle$ given
that at $\tau = 0$, the atom was in the state $|3\rangle
[|4\rangle]$. In the limit of large $\tau$,
\begin{equation}
G_{\pi}^{(2)}(\tau)
\rightarrow(\frac{\omega_{0}}{c})^{8}\frac{|\mathcal{D}|^4}{36r^4}
(\frac{2|\Omega_{c}|^2}{\lbrack2|\Omega_{c}|^2+\Gamma^2+\Delta^2\rbrack})\;,
\end{equation}
 Next let us derive the expression for two-time
photon-photon correlation in absence of interference. In this case
the total photon-photon correlation will be a simple addition of
photon-photon correlations for radiation emitted on individual
$\pi$ transitions.
\begin{eqnarray}
\label{eq1} \mathsf{G}_{\pi}^{(2)}(\tau)& = &\langle
I_{\pi}(t+\tau)I_{\pi}(t) \rangle\nonumber\\& =&
\langle\mathbf{E}^{-}_{\pi}(\vec{r},t)\mathbf{E}^{-}_{\pi}(\vec{r},t+\tau):\;\nonumber\\
&&\mathbf{E}^{+}_{\pi}(\vec{r},t+\tau)\mathbf{E}^{+}_{\pi}(\vec{r},t)\rangle_{|1\rangle\langle
3|}\;\nonumber\\
&+&\langle\mathbf{E}^{-}_{\pi}(\vec{r},t)\mathbf{E}^{-}_{\pi}(\vec{r},t+\tau):\;\nonumber\\
&&\mathbf{E}^{+}_{\pi}(\vec{r},t+\tau)\mathbf{E}^{+}_{\pi}(\vec{r},t)\rangle_{|2\rangle\langle
4|}\;,\nonumber\\
\\
& = &(\frac{\omega_{0}}{c})^{8}\frac{1}{r^4}
\{[\hat{n}\times(\hat{n}\times\vec{d}_{31})]^{\ast}\nonumber\\
&&\cdot[\hat{n}\times(\hat{n}\times\vec{d}_{31})]\}^{2}\nonumber\\
&&\langle|1\rangle\langle 3|_{t}(|1\rangle\langle
1|)_{t+\tau}|3\rangle\langle 1|_{t}\rangle\nonumber\\
&+&(\frac{\omega_{0}}{c})^{8}\frac{1}{r^4}
\{[\hat{n}\times(\hat{n}\times\vec{d}_{42})]^{\ast}\nonumber\\
&&\cdot[\hat{n}\times(\hat{n}\times\vec{d}_{42})]\}^{2}\nonumber\\
&&\langle|2\rangle\langle 4|_{t}(|2\rangle\langle
2|)_{t+\tau}|4\rangle\langle 2|_{t}\rangle\;,
\end{eqnarray}
Finally using Eq. (21),(22) and (12) we get the photon-photon
correlation in absence of interference as
\begin{eqnarray}
\mathsf{G}_{\pi}^{(2)}(\tau)&
=&(\frac{\omega_{0}}{c})^{8}\frac{|\mathcal{D}|^4}{36r^4}(f_{12}(\tau)
+f_{56}(\tau))\nonumber\\
&\times&(\frac{1}{2}\frac{|\Omega_{c}|^2}{\lbrack2|\Omega_{c}|^2+\Gamma^2+\Delta^2\rbrack})\;,
\end{eqnarray}
Here $f_{12}(\tau) [f_{56}(\tau)]$ is the probability of finding
the atom in the states $|1\rangle $ [$|2\rangle$] given that at
$\tau = 0$, the atom was in the state $|3\rangle$ [$|4\rangle$].
Eq. (32) in the limit of large $\tau$ becomes,
\begin{equation}
\mathsf{G}_{\pi}^{(2)}(\tau)\rightarrow(\frac{\omega_{0}}{c})^{8}\frac{|\mathcal{D}|^4}{36r^4}
(\frac{|\Omega_{c}|^2}{\lbrack2|\Omega_{c}|^2+\Gamma^2+\Delta^2\rbrack})\;,
\end{equation}
We now further calculate the normalized photon-photon correlation
corresponding to Eq. (28) and Eq. (32). The $g^{(2)}$ function
gives the non-classical aspects of photon statistics.
\begin{widetext}
\begin{eqnarray}
g^{(2)}(t+\tau,t) = \frac{\langle I_{\pi}(t+\tau)I_{\pi}(t)
\rangle}{\langle I_{\pi}(t+\tau)\rangle \langle I_{\pi}(t)
\rangle}=\frac{(F_{2}(\tau)+F_{6}(\tau))\tilde{\rho}_{11}}{4\tilde{\rho}^2_{11}}\;,
\end{eqnarray}
\end{widetext}
\begin{widetext}
\begin{eqnarray}
\mathsf{g}^{(2)}(t+\tau,t)& = & \frac{\langle I_{\pi}(t+\tau)
I_{\pi}(t) \rangle}{(\langle I_{\pi}(t+\tau)\rangle \langle
I_{\pi}(t) \rangle)_{|1\rangle\langle 3|}+ (\langle
I_{\pi}(t+\tau)\rangle \langle I_{\pi}(t)
\rangle)_{|2\rangle\langle 4|}} =
\frac{(f_{12}(\tau)+f_{56}(\tau))\tilde{\rho}_{11}}{2\tilde{\rho}^2_{11}}\;,
\end{eqnarray}
\end{widetext}
Here $\tilde{\rho}_{11}$ is the steady state population of the
excited state given by Eq. (12) and $g^{(2)}$ [$\mathsf{g}^{(2)}$]
is the normalized two time photon-photon correlation function
corresponding to presence [absence] of vacuum induced
interference.

\section{Numerical Results}
In this section we present our numerical results and discuss their
consequences. To begin with, we first discuss our method of
computation. The decay rates of the excited states to the two
ground states, $2\gamma_{\sigma}$ and $2\gamma$ are proportional
to $|\vec{d}_{41}|^2$ and $|\vec{d}_{31}|^2$ respectively. From
Eq. (2) we get, $2\gamma_{\sigma} \equiv\gamma_{0}/3$ and $2\gamma
\equiv\gamma_{0}/6$ , where $\gamma_{0}$ is proportional to the
square of the reduced dipole matrix element. We use these values
for the decays in our numerical computation and normalize all the
computational parameters with respect to $\gamma_{0}$. Further we
use standard subroutines to diagonalize the complex general matrix
$\mathcal{M}$ and obtain complex eigenvalues and eigenvectors of
the form $(\alpha+i\beta).$
\begin{table}[!h]
\caption{Eigenvalues for the diagonalized matrix $\mathcal{M}$
corresponding to two different values of the Rabi frequency of the
driving field which is on resonance with the atomic transitions.}
\begin{tabular}{|c|c|c|}

    \hline
$\lambda$ &  $\Omega = 0.5\gamma_{0}$    &   $\Omega = 3.0\gamma_{0}$   \\
    \hline
1   & (-0.349797,-1.10904) & (-0.375000,5.99870)  \\
2   & (-0.349797,1.10904) & (-0.375000,-5.99870) \\
3   & (-0.215794,-1.09726) & (-0.208269,5.99522) \\
4   & (-0.215794,1.09726) & (-0.208269,-5.99522)  \\
5   & (-0.300406,0.000000) & (-0.250000,0.000000)\\
6   & (-0.165314,0.000000) &(-0.250000,0.000000)\\
7   & (-0.403098,0.000000) & (-0.333462,0.000000)\\
8   & (0.000000,0.000000) & (0.000000,0.000000) \\
\hline
\end{tabular}
\end{table}
\begin{figure}[!h]
\begin{center}
\scalebox{0.55}{\includegraphics{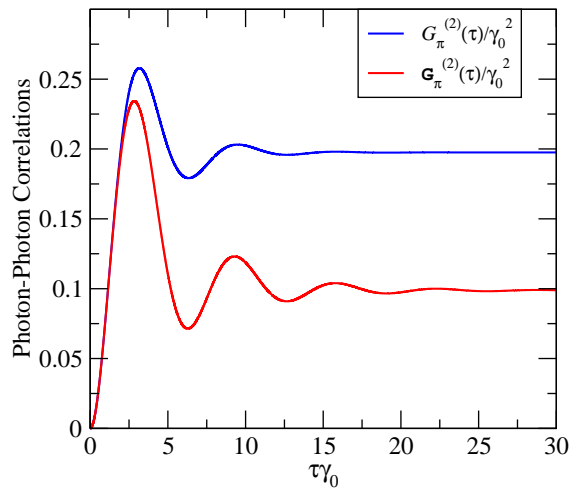}} \caption{Plot of
two-time Photon-Photon correlation as a function of time for
$\Omega = 0.5,\Delta = 0.0$. All the parameters are normalized
with respect to $\gamma_{0}$, where $\gamma_{0} =
\frac{4|\mathcal{D}|^2\omega^3_{14}}{3c^3}$ The blue and red lines
correspond to photon-photon correlations in presence and absence
of VIC respectively. }
\end{center}
\end{figure}
\begin{figure}[!h]
\begin{center}
\scalebox{0.55}{\includegraphics{fig2.eps}} \caption{Plot of
two-time Photon-Photon correlation as a function of time but now
for a small detuning $\Delta = 0.5$. other parameters remaining
same as in Fig.2}
\end{center}
\end{figure}
\begin{figure}[!h]
\begin{center}
\scalebox{0.60}{\includegraphics{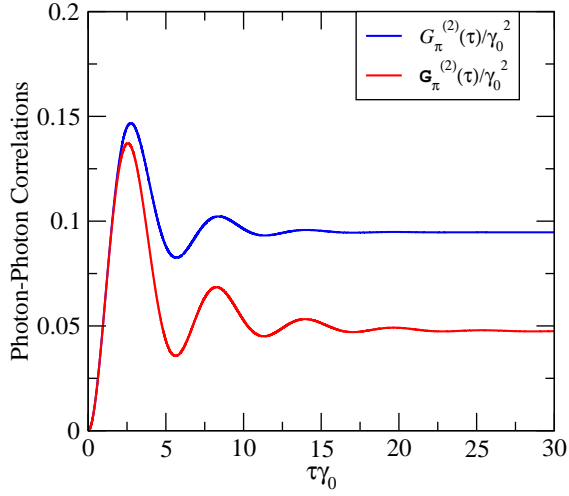}} \caption{Plot of
two-time Photon-Photon correlation as a function of time for
$\Omega = 3.0,\Delta = 0.0$. All the parameters are normalized
with respect to $\gamma_{0}$, where $\gamma_{0} =
\frac{4|\mathcal{D}|^2\omega^3_{14}}{3c^3}$. }
\end{center}
\end{figure}
\begin{figure}[!h]
\begin{center}
\scalebox{0.55}{\includegraphics{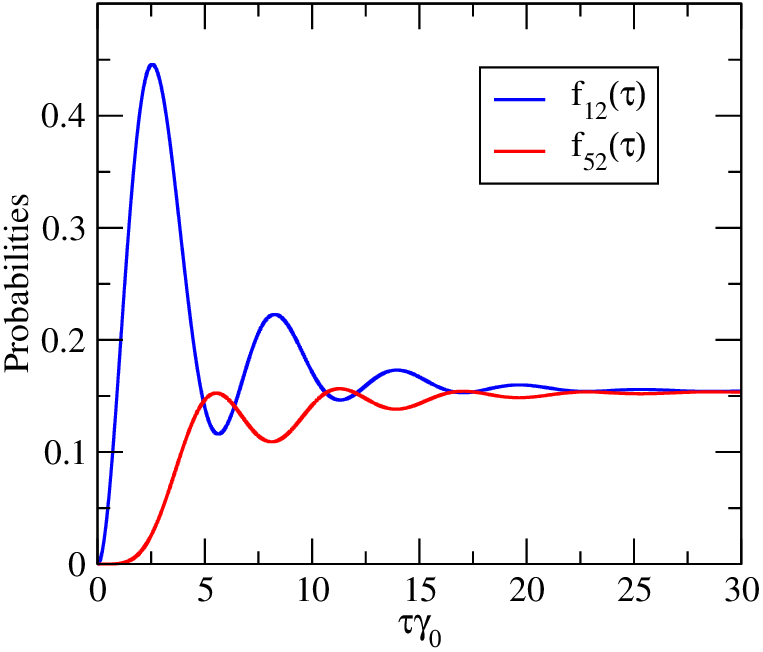}} \caption{ Probability
for finding the atom in state $|1\rangle$ $(f_{12})$ and
$|2\rangle$ $(f_{52})$ given that at time $\tau = 0$ the atom was
in the state $|3\rangle$ for $\Omega = 0.5,\Delta = 0.5$. All the
parameters are normalized with respect to $\gamma_{0}$,
$\gamma_{0} = \frac{4|\mathcal{D}|^2\omega^3_{14}}{3c^3}$.}
\end{center}
\end{figure}
\begin{figure}[!h]
\begin{center}
\scalebox{0.60}{\includegraphics{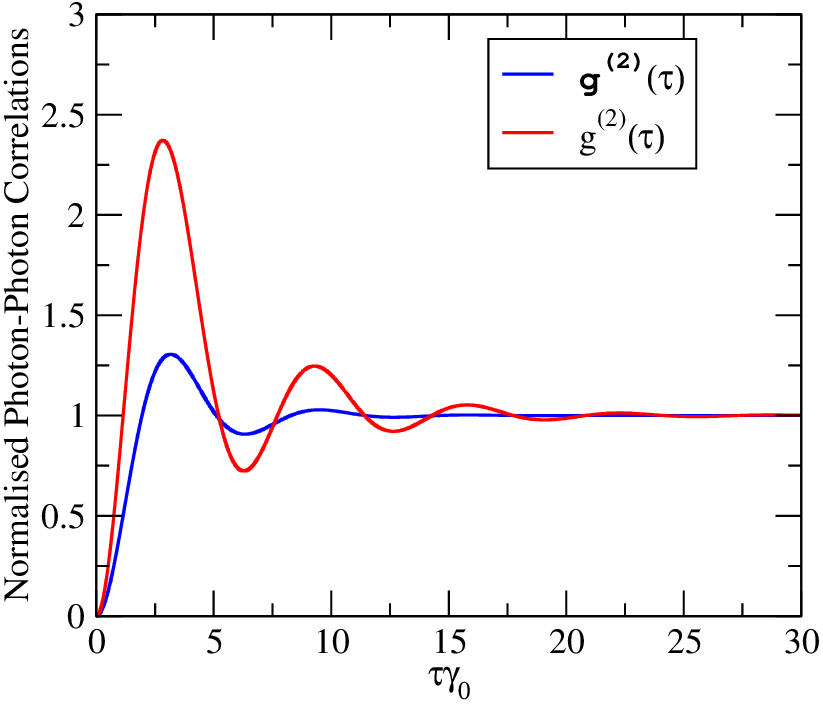}} \caption{ Normalized
photon-photon correlations plotted as a function of time, for
$\Omega = 0.5,\Delta = 0.0$. All the parameters are normalized
with respect to $\gamma_{0}$, where $\gamma_{0} =
\frac{4|\mathcal{D}|^2\omega^3_{14}}{3c^3}$.}
\end{center}
\end{figure}
For all values of detuning and Rabi frequency used in our
computation we have two pairs of complex conjugate eigenvalues and
four other eigenvalues whose complex part are so small compare to
the real part that these complex parts have no significant
contributions. Hence these four eigenvalues can be taken to be
purely real. Note that this is in contrast to the case of
photon-photon correlations for the two level model where the
number of eigenvalues is four \cite{mo}. The changes in the
eigenvalues lead to spectral modification as discussed by Kiffner.
et. al. \cite{evers}. The eigenvalues for $\Omega_{c} =
0.5\gamma_{0}$, and $\Omega_{c}=3\gamma_{0}$ and detuning $\Delta
= 0 $ are listed in the Table (I). Note for example that for
$\Omega_{c} = 3\gamma_{0}$ we have eigenvalues $\pm
5.99870i-0.375$ and $\pm 5.99522i-0.208269$. This difference in
the real parts can produce a dip in the side bands in the Mollow
spectrum. Next we calculate the elements $f_{ij}$ of the
8$\times$8 matrix [\textit{f}] using Eq. (23) and Eq. (24).
Finally we use the elements $f_{ij}$ corresponding to Eqs.
(28),(32) and Eqs. (34),(35) to evaluate the two time
photon-photon correlations and normalized photon-photon
correlations in presence and absence of
vacuum induced interference respectively.\\
The Figs. (2-4) show photon-photon correlations corresponding to
Eqs. (28) and (32). The blue and red line in the figures
correspond respectively, to photon-photon correlations in presence
and absence of interference. The correlations calculated in
presence of interference show strong damping of the oscillations
and attain an overall higher value as the time separation $\tau$
between two counts increases. The differences between $G^{(2)}$
and $\mathsf{G}^{(2)}$ are most noticeable in the limit of large
time separation $\tau$. In order to understand this we examine the
distinction between $F_{2}(\tau)=f_{12}(\tau)+f_{52}(\tau)$ and
$f_{12}(\tau)$. We recall that $f_{12} [f_{52}]$ was the
probability of finding the atom in the state $|1\rangle
[|2\rangle]$ given that at $\tau=0$, it was in the state
$|3\rangle$. We exhibit these probabilities in the Fig. (5). We
observe that the function $f_{52}(\tau)$ starts becoming
significant at the time scale of the order of
$\gamma^{-1}_{\sigma}$.\\
Further for large $\tau$, $f_{12}$ and $f_{52}$ become comparable.
The physical process that contributes to $f_{52}$ is the
following,
\begin{equation}
\xymatrix{|3\rangle \ar[r]^{laser}_{\pi-pol} &
|1\rangle
 \ar[r]^{\sigma-photon}_{emission} & |4\rangle
 \ar[r]^{laser}_{\pi-pol} & |2\rangle}.\;\nonumber
\end{equation}
Similarly population can start from the state $|4\rangle$ and end
up in the state $|1\rangle$ via,
\begin{equation}
\xymatrix{|4\rangle \ar[r]^{laser}_{\pi-pol} & |2\rangle
 \ar[r]^{\sigma-photon}_{emission} & |3\rangle
 \ar[r]^{laser}_{\pi-pol} & |1\rangle}.\;\nonumber
\end{equation}
We show normalized photon-photon correlations in a typical case in
the Fig. (6). In case of interference we observe stronger damping
of the oscillations and an overall reduction of the $g^{(2)}$
function at shorter time scales. At long time limits
$g^{(2)}(\tau\rightarrow\infty)$ is 1.
 Photon  antibunching  effect is also visible as $0 \leqslant g^{(2)}(0)<1$.
 For shorter time scale we get $g^{(2)}(\tau)\not\le 1$ a clear
 signature of the nonclassical nature of the two-time
 photon-photon correlations.

\section{Conclusions}
In conclusion we have shown that the vacuum induced coherence(VIC)
do significantly affect the two-time photon-photon correlations
even though they show no effect on the total steady state
intensity of the radiation emitted on the $\pi$ transitions. The
effect of this coherence is reflected in form of stronger damping
and overall larger values of the correlation function $G^{(2)}$.
The level scheme j = 1/2 $\rightarrow$ j = 1/2 is easily
realizable and has already been used, for example in
$^{198}Hg^{+}$ \cite{eich} in the context of interferences
produced by a system of two ions and more recently in
$^{138}Ba^{+}$ \cite{blatt} in the context of emission in presence
of a mirror. In future we hope to investigate how the asymmetry in
the level structures introduced by a magnetic field \cite{sunish}
would influence the photon-photon correlations. This might in turn
give us more freedom in choosing the level structure and hence
more broader choice in selecting atomic transition for
experiments. Finally note that it would also be interesting to
examine the VIC effects in the context of nonlinear optical
experiments using j=1/2 to j=1/2 transitions.

\end{document}